\documentclass[pra,amsmath,amssymb,twocolumn,10pt,showpacs]{revtex4}
\usepackage{color}
\usepackage{braket}
\usepackage{enumitem}
\usepackage{pifont}

\DeclareSymbolFont{symbolsC}{U}{txsyc}{m}{n}
\DeclareMathSymbol{\strictif}{\mathrel}{symbolsC}{74}
\DeclareMathSymbol{\boxright}{\mathrel}{symbolsC}{128}

\newcommand{\blk}{\color{black}}

\newcommand{\beq}{\begin{equation}}
\newcommand{\eeq}{\end{equation}}
\newcommand{\nn}{\nonumber}

\newcommand{\dg}{^\dagger}

\newcommand{\cu}[1]{\left\{ {#1} \right\}}

\newtheorem{theorem}{Theorem}

\newtheorem{corollary}[theorem]{Corollary}

\newtheorem{definition}[theorem]{Definition}

\begin{document}

\title{Unavoidable decoherence in the quantum control of an unknown state}

\author{D. Kielpinski${}^{1,2}$, R. A. Briggs${}^{2,3}$, and H. M. Wiseman${}^{1,2,4}$}

\affiliation{${}^1$Centre for Quantum Dynamics and ${}^2$School of Biomolecular and Physical Sciences, Griffith University, Nathan QLD 4111, Australia \\
${}^3$ School of Philosophy, Australian National University, Canberra ACT 0200, Australia \\
${}^4$ ARC Centre of Excellence for Quantum Computation and Communication Technology, Griffith University, Nathan QLD 4111, Australia}

\begin{abstract}

A common objective for quantum control is to force a quantum system, initially in an unknown state, into a particular target subspace.  We show that if the subspace is required to be a decoherence-free subspace of dimension greater than 1, then such control must be decoherent. That is, it will take almost any pure state to a mixed state. We make no assumptions about the control mechanism, but our result implies that for this purpose coherent control offers no advantage, in principle, over the obvious measurement-based feedback protocol. 
\end{abstract}
\pacs{quantum decoherence, quantum measurement, quantum control, decoherence-free subspace}
\maketitle

\section{Introduction}

The application of control theory techniques to quantum systems promises technological improvements in a great variety of areas \cite{Wiseman-Milburn-quantum-control-BOOK}, including quantum information processing \cite{Khodjasteh-Viola-decoherence-free-dynamical-control, Kerckhoff-Mabuchi-coherent-feedback-error-correction, Santos-Carvalho-quantum-jump-QC, Vijay-Siddiqi-SC-rabi-oscillation-feedback}, sub-shot-noise metrology \cite{Xiang-Pryde-unknown-phase-entangled-measurement, Yonezawa-Furusawa-quantum-enhanced-phase-tracking}, creating non-classical states \cite{Sayrin-Haroche-photon-number-feedback-preparation, Norris-Deutsch-initial-state-squeezing-enhancement}, efficient state-tracking \cite{Karasik-Wiseman-efficient-state-tracking}, and chemical analysis \cite{Roth-Rabitz-quantum-control-chemical-analysis}. The field of quantum control theory studies the strategies that a controller can use to drive a quantum system to attain some desirable properties, referred to as the target. Like classical control, quantum control can be classified according to the types of operations available to the controller \cite{Wiseman-Milburn-quantum-control-BOOK}. In open-loop quantum control, the  
controller induces unconditional quantum dynamics leading to the target; if only unitary dynamics are used, one must know the relevant properties of the initial state for the control to be successful. {So-called} learning quantum control is a way of refining open-loop quantum control for complicated systems, using measurement of the final state of the system to judge the success of the control strategy and modify the strategy on the next trial.  In measurement-based quantum feedback, the system state is typically measured while the control is being applied, and the results used to set the control, so that even unknown states of the system can be driven towards the target. Coherent feedback control is similar, but the controller is allowed to be fully quantum mechanical, with no measurement step assumed \cite{Wiseman-Milburn-coherent-quantum-feedback, Lloyd-coherent-quantum-feedback}.

Quantum control will, in general, cause decoherence. Even if the controls applied are unitary, the measurement step in measurement-based
feedback, or the presence of an auxiliary sytem in coherent quantum feedback,  generally takes an initially pure state to a mixed final state. In this paper we show that such decoherence is inevitable for the very natural control objective of forcing the system into a decoherence-free subspace (DFS) \cite{Zanardi-Rasetti-DFS, Lidar-Whaley-DFS}. The DFS restriction means that if the system is already in the desired subspace, it should not be disturbed. We note that if $A$ denotes the property of being in the desired subspace, then this restriction is also frequently proposed in causal decision theory as a constraint on any operation that can be considered to make $A$ hold \cite{Pearl-causality-BOOK, Skyrms-Causal-Factors-INBOOK, Gibbard-Two-Recent-Theories-INCOLLECTION, Stalnaker-Theory-of-Conditionals-INCOLLECTION}.  Moreover, we show that any quantum control of this type can always be realised by a projective measurement to determine whether $A$ initially holds, followed by corrective action if it is found not to, a procedure that can be thought of as a one-step measurement-based feedback. Our proof characterises all possible evolutions---all possible trace-preserving completely positive (TPCP) maps---that implement such strategies. Since TPCP maps are the most general allowed operations of quantum dynamics, our result is independent of the common assumption of Markovian control dynamics, which was recently used in proving results related to ours \cite{Bolognani-Ticozzi-markovian-stability-subspaces, Milburn-decoherence-classical-control-conditions}.

\section{Result}
We denote the set of bounded operators on some Hilbert space $\mathsf{H}$ by ${\mathfrak B}(\mathsf{H})$, while ${\mathfrak U}(\mathsf{H})$ similarly denotes the set of unitary operators. The decoherence-free condition on our control map is then stated formally as follows \cite{Zanardi-Rasetti-DFS, Lidar-Whaley-DFS}:
\begin{definition} 
A subspace $\mathsf{H}_A \subseteq \mathsf{H}$ is said to be decoherence-free under a TPCP map ${\cal S}$ from ${\mathfrak B}(\mathsf{H})$ to ${\mathfrak B}(\mathsf{H})$ iff
\begin{equation}
\exists\ U_A \in {\mathfrak U}(\mathsf{H}_A):\ \forall \rho_A \in {\mathfrak B}(\mathsf{H}_A), \ 
{\cal S} (\rho_A)  = U_A \rho U_A\dg
\end{equation}
\end{definition}

Our result is now formally stated as
\begin{theorem}
Let ${\cal S}$ be a TPCP map from ${\mathfrak B}(\mathsf{H})$ to ${\mathfrak B}(\mathsf{H}_A)$ with $\mathsf{H} = \mathsf{H}_A \oplus \mathsf{H}_{\bar{A}}$. That is,
\beq
{\cal P}_A{\cal S} = {\cal S}, \label{effective}
\eeq
where ${\cal P}_A$ is the projection superoperator onto ${\mathsf H}_A$.

Then if $\mathsf{H}_A$ is a DFS under ${\cal S}$ it follows that \beq {\cal S} = {\cal S} \left( {\cal P}_A + {\cal P}_{\bar{A}} \right). \label{result}
\eeq
\end{theorem}
That is, the map necessarily destroys coherence between the $A$ and $\bar{A}$ subspaces. Since
\beq
{\cal S} {\cal P}_A = {\cal U}_A {\cal P}_A,  \label{dfs}
\eeq
where ${\cal U}_A(\bullet)= U_A \bullet U_A\dg$, the significance of this result is the following corollary:
\begin{corollary}
Let $\mathsf{H}_A$ be a subspace of $\mathsf{H}$ with $1 < \dim \mathsf{H}_A < \dim \mathsf{H}$. Consider an arbitrary $\ket{\psi} \in \mathsf{H}$, which can be written $\ket{\psi} = \ket{\psi_A} \oplus \ket{\psi_{\bar{A}}}$. Then under the conditions of the theorem, ${\cal S}(\ket{\psi}\bra{\psi})$ is mixed except for the set of measure zero where $${\cal S}(\ket{\psi_{\bar{A}}}\bra{\psi_{\bar{A}}}) \propto U_A \ket{\psi_A}\bra{\psi_A} U_A\dg.$$
\end{corollary}

That is, almost all initial pure states will become mixed under any map which satisfies the conditions of the theorem, and which acts nontrivially ($\mathsf{H}_A \neq \mathsf{H}$) and for which the DFS carries a nonzero amount of information, equal to $\log_2(\dim \mathsf{H}_A)$ qubits to be specific.

\section{Proof}

\blk
To prove the result, we work in the matrix block representation of $\mathsf{H} = \mathsf{H}_A \oplus \mathsf{H}_{\bar{A}}$ and write the {initial density matrix} $\rho$ as
\begin{align}
\rho &= \left( \begin{array}{c|c} \rho_A & \rho_C \\ \hline \rho_C^\dagger & \rho_{\bar{A}} \end{array} \right),
\end{align}
where the $C$ subscript is used because $\rho_C$ represents the initial coherences between $A$ and $\bar{A}$. Decomposing ${\cal S}$ into its Kraus operators $K_\mu$, i.e.
\beq
{\cal S}(\rho) \equiv \sum_\mu K_\mu \rho K_\mu^\dagger,
\eeq
condition (\ref{effective}) implies that there exist some operators $A_\mu, B_\mu$ such that
\begin{align}
K_\mu &= \left( \begin{array}{c|c} A_\mu & B_\mu \\ \hline 0 & 0 \end{array} \right)
\end{align}
This ensures that
\begin{align}
{\cal S}(\rho) &= \sum_\mu \left( \begin{array}{c|c} D_\mu & 0 \\ \hline 0 & 0 \end{array} \right), \textrm{ where } \\
D_\mu  & \blk\equiv A_\mu \rho_A A_\mu^\dagger + B_\mu \rho_{\bar{A}} B_\mu^\dagger + \left( A_\mu \rho_C B_\mu^\dagger + \mbox{h.c.} \right)
\end{align}
Since ${\cal S}$ is trace-preserving, we have
\begin{align}
\sum_\mu K_\mu^\dagger K_\mu &= \sum_\mu \left( \begin{array}{c|c} A_\mu^\dagger A_\mu & A_\mu^\dagger B_\mu \\
\hline B_\mu^\dagger A_\mu & B_\mu^\dagger B_\mu \end{array} \right) = I , \label{tpcpmat}
\end{align}
which implies that $\sum_\mu A_\mu^\dagger B_\mu = 0$.

Now by condition (\ref{dfs}), there exists a unitary operator $U_A$ such that $A_\mu \propto U_A$ for all $\mu$, so
\begin{align}
\sum_\mu A_\mu \rho_C B_\mu^\dagger &= U_A \rho_C U_A^{-1} \sum_\mu A_\mu B_\mu^\dagger = 0
\end{align}
We are left with
\begin{align}
D_\mu &= A_\mu \rho_A A_\mu^\dagger + B_\mu \rho_{\bar{A}} B_\mu^\dagger .
\end{align}
Thus for any $\rho$,
\begin{align}
{\cal S} \left( [{\cal P}_A + {\cal P}_{\bar{A}}]\rho \right)  &= \sum_\mu \left[ A_\mu {\cal P}_A(\rho) A_\mu^\dagger +
B_\mu {\cal P}_{\bar{A}}(\rho)  B_\mu^\dagger \right]  \nn \\
&= \sum_\mu \left( A_\mu \rho_A A_\mu^\dagger + B_\mu \rho_{\bar{A}} B_\mu^\dagger \right) \nn \\
&= {\cal S}\rho,
\end{align}
which proves the theorem.

\section{Discussion}

The above result shows that any map that satisfies the conditions of the theorem can be achieved using a simple measurement-based control as follows. First make a measurement that projects the system into the $A$ subspace or the complementary $\bar{A}$ subspace. The measurement need not be back-action evading, but its action of the $A$ subspace following projection must be unitary, described by $U_A$ (which can of course be the identity). If the $A$ result is obtain, do nothing further. If the result $\bar{A}$ is obtained, implement a TPCP map described by Kraus operators $\cu{ B_\mu }$ which map from ${\mathsf H}_{\bar{A}}$ to ${\mathsf H}_A$. If $\dim {\mathsf H}_{\bar{A}} > \dim {\mathsf H}_A$ then more than one such Kraus operator will be needed.

When the target subspace is not required to be a DFS, the above result does not hold, and there are control strategies that force the system in the desired subspace that are not equivalent to the measurement-based feedback strategy described above. For instance, consider a two-qubit system with basis states $\{ \ket{ij} \}$ with $i, j = 0,1$ labeling the state of each qubit. Let $\mathsf{H}_A$ be the subspace spanned by $\{ \ket{00}, \ket{01} \}$, for which qubit 1 is in state $\ket{0}$. Now consider ${\cal S}$ defined by the $4 \times 4$-dimensional Kraus operators
\begin{align}
K_1 &= \frac{1}{\sqrt{2}} \left( \begin{array}{c|c} I & Z \\ \hline 0 & 0 \end{array} \right) \\
K_2 &= \frac{1}{\sqrt{2}} \left( \begin{array}{c|c} Z & -I \\ \hline 0 & 0 \end{array} \right), \textrm{ where }\\
I &\equiv \left( \begin{array}{cc} 1 & 0 \\ 0 & 1 \end{array} \right)  \textrm{  and } Z \equiv \left( \begin{array}{cc} 1 & 0 \\ 0 & -1 \end{array} \right). \nn
\end{align}
Clearly condition (\ref{effective}) holds for ${\cal S}$. Since Eq. (\ref{tpcpmat}) is satisfied, ${\cal S}$ is a TPCP map. In this representation, 
\begin{align}
\sum_\mu A_\mu \rho_C B_\mu^\dagger &= [\rho_C, Z]/2
\end{align}
so that
\begin{align}
S \rho &= A_\mu \rho_A A_\mu^\dagger + B_\mu \rho_{\bar{A}} B_\mu^\dagger + [\rho_C, Z].
\end{align}
That is, the final state depends on the coherence between $\mathsf{H}_A$ and $\mathsf{H}_{\bar{A}}$ and ${\cal S}$ is inequivalent to the measurement-based feedback described above.

We have shown that all quantum control strategies that force  an {initially} unknown state into a decoherence-free subspace can be reduced to a simple strategy using projective measurement followed by conditional operations. In particular, coherent quantum feedback provides no advantage in these cases. Our proof avoids model-dependent assumptions, in particular the assumption of Markovian dynamics. It is obvious that decoherence necessarily arises in controlling an unknown quantum state into a target subspace: our result shows how this decoherence must be apportioned between the target subspace and the initial state coherence in the limit of a decoherence-free target. Future work could explore the converse: how far must the decoherence-free property of the subspace be violated in order to preserve the coherence between desirable and undesirable components of the initial state? Alternatively, one could investigate the maximal decoherence-free subspace of a given control strategy, quantifying the prior knowledge of the initial state that would guarantee fully coherent control.

This work was supported by the Australian Research Council under FT110100513 (DK, Future Fellowship) and CE110001027 (HMW, Centre of Excellence), and by the ARO MURI grant W911NF-11-1-0268 (HMW).


\end{document}